# MolCAP: Molecular Chemical reActivity pretraining and prompted-finetuning enhanced molecular representation learning

Yu Wang[1,2#], JingJie Zhang[1,2#], Junru Jin[1,2], and Leyi Wei[1,2*]

[1] School of Software, Shandong University, Jinan 250101, China

[2] Joint SDU-NTU Centre for Artificial Intelligence Research (C-FAIR), Shandong University, Jinan 250101, China

[#]These authors contributed equally to this work as the first authors

*Corresponding author:

Leyi Wei: weileyi@sdu.edu.cn

## Abstract

Molecular representation learning (MRL) is a fundamental task for drug discovery. However, previous deep-learning (DL) methods focus excessively on learning robust inner-molecular representations by mask-dominated pretraining framework, neglecting abundant chemical reactivity molecular relationships that have been demonstrated as the determining factor for various molecular property prediction tasks. Here, we present MolCAP to promote MRL, a graph pretraining Transformer based on chemical reactivity (IMR) knowledge with prompted finetuning. Results show that MolCAP outperforms comparative methods based on traditional molecular pretraining framework, in 13 publicly available molecular datasets across a diversity of biomedical tasks. Prompted by MolCAP, even basic graph neural networks are capable of achieving surprising performance that outperforms previous models, indicating the promising prospect of applying reactivity information for MRL. In addition, manual designed molecular templets are potential to uncover the dataset bias. All in all, we expect our MolCAP to gain more chemical meaningful insights for the entire process of drug discovery.

## Introduction

Molecular representation learning (MRL) is a critical process for translating molecules to a real vector space, understandable by computational algorithms. This approach is essential for in-silico drug discovery, which promises to revolutionize pharmaceutical drug development. As traditional drug development methods tend to be both time-consuming and expensive, virtual drug discovery presents an attractive alternative. Only a small fraction of drug candidates (~10%) are successfully approved after the clinical phases[1,2]., hence the need for efficient and reliable in-silico approaches.

Over the past few decades, the field of deep learning-based molecular representation learning (MRL) has experienced significant growth. The use of informative and knowledgeable molecular representations has proven to be beneficial in various drug-related tasks, including molecular property predictions[3-5], protein-ligand design[6,7], drug-drug responses[8,9], and chemical reaction predictions[10-12]. With the exponential growth in the availability of chemical experimental data, large-scale self-supervised learning (SSL) for molecular representation approaches is rapidly emerging[3,13-15]. However, most of these approaches primarily focus on the intrinsic information of the molecular structures without any extra chemical rule-based knowledge. For instance, Wang et al.[16] introduced three contrastive-based SSL tasks on molecular topologies for pretraining, while Fang et al.[17] pretrained conformation and bond-angle graph models. In addition to the aforementioned graph-based SSL approaches, SSL frameworks based on sequences[18] and images[19] have also been developed, with a focus on character-level and pixel-level perspectives, respectively. Despite these advancements, the vast chemical space poses a challenge for these models, as they heavily rely on the quality of augmented data and may struggle to generalize to different tasks. To address this issue, it is crucial to blend chemical knowledge to guide models to understand the chemical meaning during the pretraining phase.

The effective integration of chemical knowledge is a crucial aspect of molecular-related tasks. To this end, researchers have incorporated various types of chemical or biomedical information into their pretraining frameworks. These frameworks include element-wise knowledge graphs[20], molecular multi-property[18,21], and heterogeneous biomedical networks[22]. However, despite the importance of chemical reactivities, few works have focused on them in the pretraining stage. This is a significant gap, as chemical reactivities are critical to a wide range of molecular-related tasks. While the previous representative reaction-aware strategies, MolR[23], has been proposed previously, suffer from limitations in leveraging reaction data. MolR, for instance, forces equivalence between the embedded vector of reactants and products, neglecting vital directional information of chemical reactions. Therefore, there is a need for more effective strategies that can leverage reaction data to improve molecular-related tasks.

Another area that has not received adequate attention is the significant distinctions between pre-training tasks and downstream tasks. Simply using pre-trained representations for downstream tasks may not lead to optimal performance[20,24]. Prompt learning has recently become a viable means of closing the divide between pre-training and fine-tuning in addition to achieving outstanding results across various natural language processing tasks[25-27]. However, while previous attempts have been made to extend prompt learning to the graph domain, such as functional prompt[20], which incorporates functional group information into molecular representations through vector addition, there still remains ample room for designing efficient prompts to narrow the gap between graph-based pre-training and the vast range of downstream tasks.

We propose the MolCAP (**mol**ecular **c**hemical **a**ctivity pretraining with **p**rompted finetuning) framework (shown in **Figure 1**) as a solution to the aforementioned issues in molecular chemical activity prompt learning for MRL. This framework integrates molecular chem reactivity information by utilizing a multi-task pretraining technique on approximately 0.7 million chemical reactions. Moreover, we introduce a molecular graph prompt paradigm that reduces the gap between pretraining and downstream tasks, enabling deep-learning models to gain insights from sensitive substructures. We have also created several manual prompt templates to enhance the similarity between pretraining tasks and specific downstream tasks, resulting in improved interpretability. To evaluate MolCAP's effectiveness against competitive baselines, we have conducted various experiments on numerous molecular property datasets. Additionally, we have scrutinized the necessity of each component of MolCAP as well as its interpretability and robustness through extensive analyses.

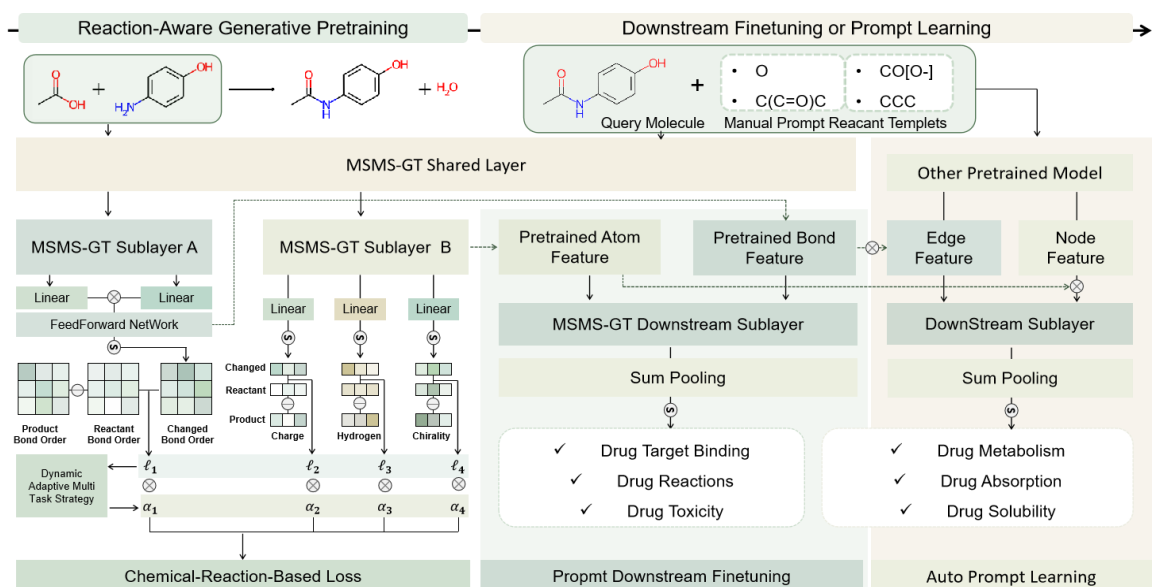

**Figure 1.** The overview of MolCAP. The pipeline of MolCAP can be categorized into generative pretraining stage and downstream finetuning stage. In the pretraining, we introduce three atom-level tasks and one bond-level task, which are scored by a multi-task learning strategy. For the finetuning stage, we propose manual prompt-based finetuning and auto-prompt finetuning. The former can bridge the gap from pretraining to finetuning; while the latter is capable of blending other models pretrained by another knowledge.

## Results

### Performance results of MolCAP on benchmark datasets

We conducte experiments on nine benchmark datasets to evaluate MolCAP's ability to predict molecular properties, including 13 public benchmarks. These datasets cover a wide range of attributes, such as target binding, drug absorption, and drug safety, and are challenging in various domains of drug discovery. The results demonstrate that MolCAP can generalize and transfer well in the biomedical field. The datasets include binary classification, multi-label classification, and regression problems, providing a comprehensive evaluation of MolCAP's predictive performance across diverse scenarios.

**Table 1-2** summarizes the comparison between MolCAP and nine other state-of-the-art approaches on the molecular property benchmarks. The baseline approach results in them were obtained from their respective original papers.

**Table 1.** Overall performance of our MolCAP and state-of-the-art methods on five benchmarks for classification tasks.

| Methods | Classification (AUC-ROC) | | | | | |
|---|---|---|---|---|---|---|
| Dataset | BACE | BBBP | ClinTox | Tox21 | SIDER | Average |
| Size | 1513 | 2039 | 1478 | 7831 | 1427 | - |
| Tasks | 1 | 1 | 2 | 12 | 27 | - |
| MPNN | 0.815 | 0.913 | 0.879 | 0.808 | 0.595 | 0.784 |
| DMPNN | 0.852 | 0.919 | 0.897 | 0.826 | 0.632 | 0.807 |
| MGCN | 0.734 | 0.850 | 0.634 | 0.707 | 0.552 | 0.690 |
| AttentiveFP | 0.863 | 0.908 | 0.933 | 0.807 | 0.605 | 0.783 |
| TrimNet | 0.843 | 0.892 | 0.906 | 0.812 | 0.606 | 0.785 |
| Mol2Vec | 0.841 | 0.876 | 0.828 | 0.805 | 0.601 | 0.774 |
| N-GRAM | 0.876 | 0.912 | 0.855 | 0.769 | 0.632 | - |
| SMILES-BERT | 0.849 | 0.959 | 0.985 | 0.803 | 0.568 | 0.803 |
| GROVER | 0.894 | 0.940 | 0.944 | 0.831 | **0.658** | 0.834 |
| MolR | 0.882 | 0.895 | 0.954 | 0.839 | - | - |
| **MolCAP (Ours)** | **0.958** | **0.961** | **0.994** | **0.840** | 0.640 | **0.879** |

Observations can be made from **Table 1-2**. Firstly, MolCAP achieves state-of-the-art performance with significant margins on most classification and regression datasets compared to previous self-supervised pretraining frameworks for molecular representation. Specifically, MolCAP has an average improvement of 12.87%, with 5.39% on common classification datasets, 20.35% on regression datasets. This indicates that MolCAP is an effective and generalized generative framework that can produce robust and transferable molecular representations. Secondly, compared with the reaction-aware pretraining framework, contrastive-learning-based MolR, there are more significant improvements for

MolCAP on the provided results. Specially, there are approximately 8.39%, 7.37%, 3.35% and 1.07% improvements on BACE, BBBP, ClinTox, and Tox21 datasets, respectively. This suggests that the multi-task generative pretraining and prompted finetuning from used in MolCAP exhibit more promising ability to utilize chemical reactivity knowledge for boosting the versatile downstream tasks than previous contrastive-learning-based pretraining strategy. Overall, these results suggest that MolCAP can be a more accurate and effective approach than existing methods for screening clinically eligible candidate compounds in the early stages of drug discovery.

**Table 2.** Overall performance of our MolCAP and state-of-the-art methods on three benchmarks for regression tasks.

| Methods | Regression (RMSE) | | | |
|---|---|---|---|---|
| Dataset | FreeSolv | ESOL | Lipo | Average |
| Size | 642 | 1128 | 4200 | - |
| Tasks | 1 | 1 | 1 | - |
| MPNN | 2.185 | 1.167 | 0.672 | 1.341 |
| DMPNN | 2.177 | 0.980 | 0.653 | 1.270 |
| MGCN | 3.349 | 1.266 | 1.113 | 1.909 |
| AttentiveFP | 2.030 | 0.853 | 0.650 | 1.178 |
| TrimNet | 2.529 | 1.282 | 0.702 | 1.504 |
| Mol2Vec | 5.752 | 2.358 | 1.178 | 3.096 |
| N-GRAM | 2.512 | 1.100 | 0.876 | 1.496 |
| SMILES-BERT | 2.974 | 0.841 | 0.666 | 1.493 |
| GROVER | 1.544 | 0.831 | **0.560** | 0.978 |
| **MolCAP (Ours)** | **0.966** | **0.691** | 0.679 | **0.779** |

**Table 3.** Overall performance of our MolCAP and comparative methods on five extended datasets for more metabolism tasks.

| Methods | Classification (AUC-ROC) | | | | | |
| --- | --- | --- | --- | --- | --- | --- |
| Dataset | CYP1A2 | CYP2C9 | CYP2C19 | CYP2D6 | CYP3A4 | Average |
| Size | 12579 | 12092 | 12665 | 13130 | 12328 | - |
| Tasks | 1 | 1 | 1 | 1 | 1 | - |
| Chemception | 0.699 | 0.733 | 0.738 | 0.753 | 0.763 | 0.737 |
| ADMET-CNN | 0.556 | 0.697 | 0.616 | 0.559 | 0.654 | 0.616 |
| QSAR-CNN | 0.712 | 0.736 | 0.752 | 0.761 | 0.734 | 0.739 |
| **MolCAP (Ours)** | **0.888** | **0.961** | **0.986** | **0.848** | **0.806** | **0.898** |

**Performance on extended larger metabolism dataset**

To estimate the expansibility of MolCAP on more specialized dataset, we compared MolCAP with three effective models on extended drug metabolism datasets. As illustrate in **Table 3**, MolCAP appears to outperform the other approaches in terms of AUC scores. Specifically, MolCAP achieves an impressive AUC-ROC score of 0.898, which is significantly higher than the other methods such as Chemception, ADMET-CNN, and QSAR-CNN. Moreover, MolCAP outperforms other methods in all five tasks, with the highest AUC-ROC score of 0.986 achieved in the CYP2C19 task. This indicates that MolCAP is a robust and effective model for predicting molecular metabolism properties across different tasks. Overall, the superior performance of MolCAP can be attributed to its activity knowledge acquired from pretraining stage. This allows MolCAP to learn more informative representations of molecules and make more accurate predictions.

**Reactivity knowledge boosts the performance using conventional methods through auto-prompt learning**

To further validate the effect of reactivity knowledge, we integrate traditional graph learning models with the reactivity information via the auto-prompt manner. To facilitate the comparition, we selected four lightweight approaches, GATV2, GATV1[28], GIN[29] and GCN[30], as anchor models. The corresponding results can be seen in **Figure 2**. As can be seen, though in ClinTox dataset slight improvement can be observed, the incorporation of reactivity knowledge significantly boosted the performance of these models in all other seven datasets, including four classification tasks: BACE, BBBP, TOX21, SIDER and three regression tasks: ESOL, FreeSolv, and Lipo. This demonstrates the importance of considering reactivity information in predicting the toxicity and bioactivity of molecules. In addition, the results also suggest that the auto-prompt method is effective in integrating domain knowledge into graph learning models.

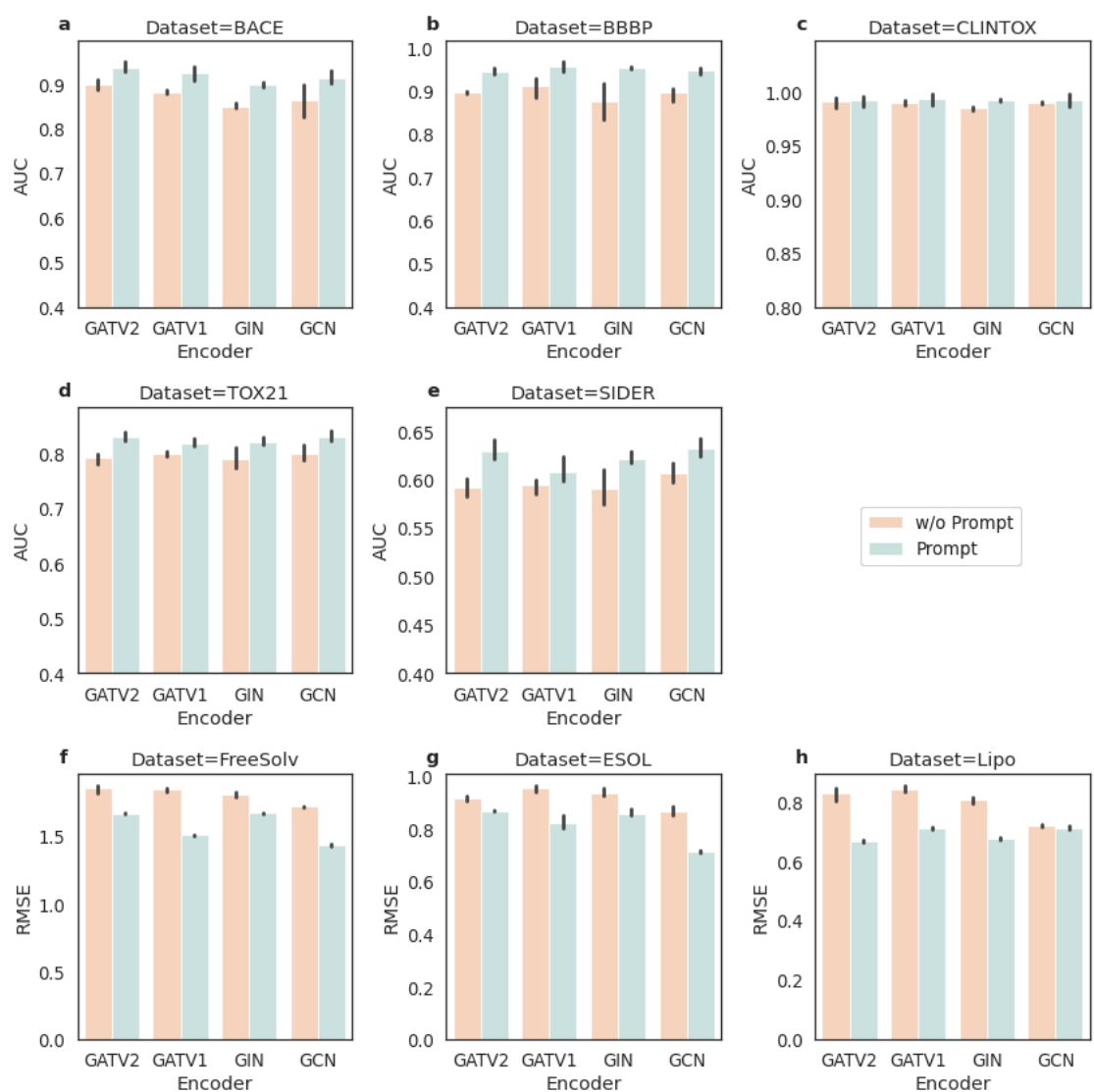

**Figure 2.** The auto-prompt results on eight biomedical datasets based on different graph learning models. a-h illustrate that promoted by chemical reactivity knowledge, various of basic graph neural networks can be promoted. a-e. illustrate the performance on classification tasks, a larger AUC score means better performance. f-h. display their RMSE results, the littler the better.

**Discover insight from manually designed prompt**

Another advantage of using chemical reaction tasks for pre-training and fine-tuning downstream tasks with artificially designed prompt templates is that different molecular templates can be used to explore the data set preferences of downstream tasks, which can promote our understanding of data set composition. Figure 3 shows the results of different molecular templates selected on four representative data sets for the final task performance. We have the following observations: 1) The existence of molecular templates has a significant improvement on the model. Among them, methyl ammonium ions, carboxylate ions, and propane molecules are the best performing templates for the SIDER, FreeSolv, and ESOL data sets, respectively. 2) Although models using molecular templates perform well on most data sets, molecular templates do not have a positive effect on all data sets. For example, in the TOX21 data set, the model without molecular templates achieved the highest AUC, 0.84. This may be because we have not yet found suitable molecular templates. 3) Looking at the categories of template molecules, some data sets are more sensitive to polar molecules or ions, such as the SIDER data set, where the model with methyl ammonium ions, carboxylate ions, and hydroxyl ions as templates generally performs better than the model with propane molecules; while in the ESOL data set, the model with non-polar propane molecules as templates has a lower RMSE error. This suggests that data sets have preferences for different molecules, namely, the SIDER data set has a preference for polar molecules, while the ESOL data set has a preference for non-polar propane molecules. In summary, the type of molecular templates may have some relationship with the performance on the test set, and this relationship implies some common rules of chemical properties of the data set.

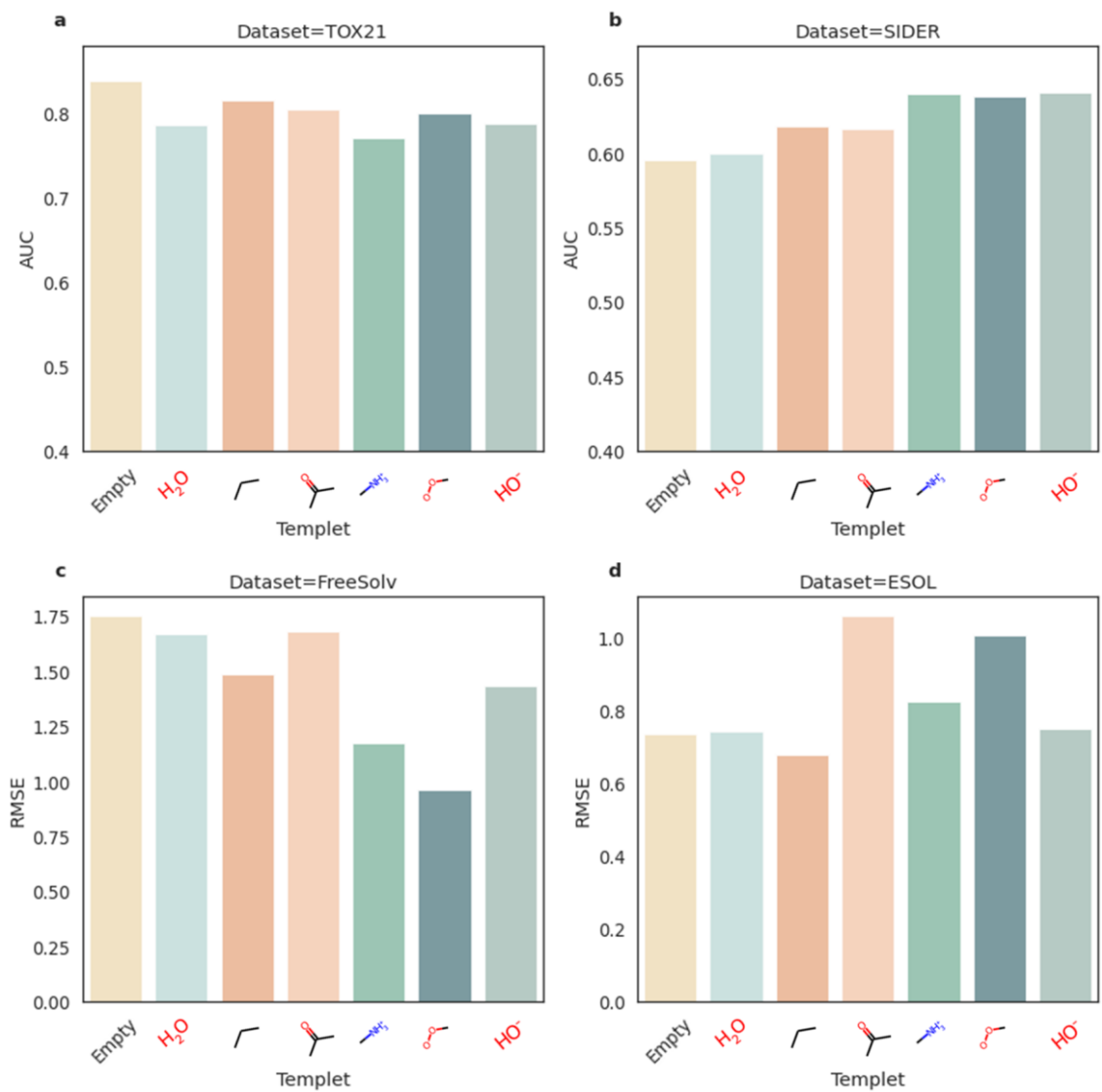

**Figure 3.** Performance using different manual prompt templets while finetuning on four representative datasets. a-b. illustrate the templet effect on two classification datasets, larger AUC scores denote better performance. c-d. display the similar results on two regression tasks, the litter the value the better the performance.

**Model Interpretabilities**

We use MolCAP to calculate atom-wise attention scores in order to explore the relationship between molecular structures and properties. This helps to bridge connections between structure and activity for a given molecule. To capture the sensitive structure of the demand property, we used global attention scores in the Graph Transformer layer to calculate aggregated atomic sensitive scores. Additionally, we defined the edge-wise sensitive score as the average of the two end atoms, which reflects the node-level contribution to the target property. Each atomic sensitive score also aggregates neighborhood information, indicating that it is aware of the contextual information.

As shown in **Figure 4**, the sensitive substructures are colored by different gradations, showing the possible quantification of structure-activity connections. For example, as for the property of BACE binding, the sensitivity score is centered on carbonyl, amidogen or double bond structures, which is highly related to the chemical reactivities. In addition, this attention-based interpretability is potential for guiding the optimization molecule design. For instance, compared with molecules in **Figure 4a**, the molecule in **Figure 4g** is classified as the positive ligand for BACE target, and the trifluoromethyl on the meso-site is considered as the key substructure by MolCAP, since these two molecular structures share the same functional groups such as amidogen, carbonyl and sulfonyl group but has two distinct properties in BACE binding conditions.

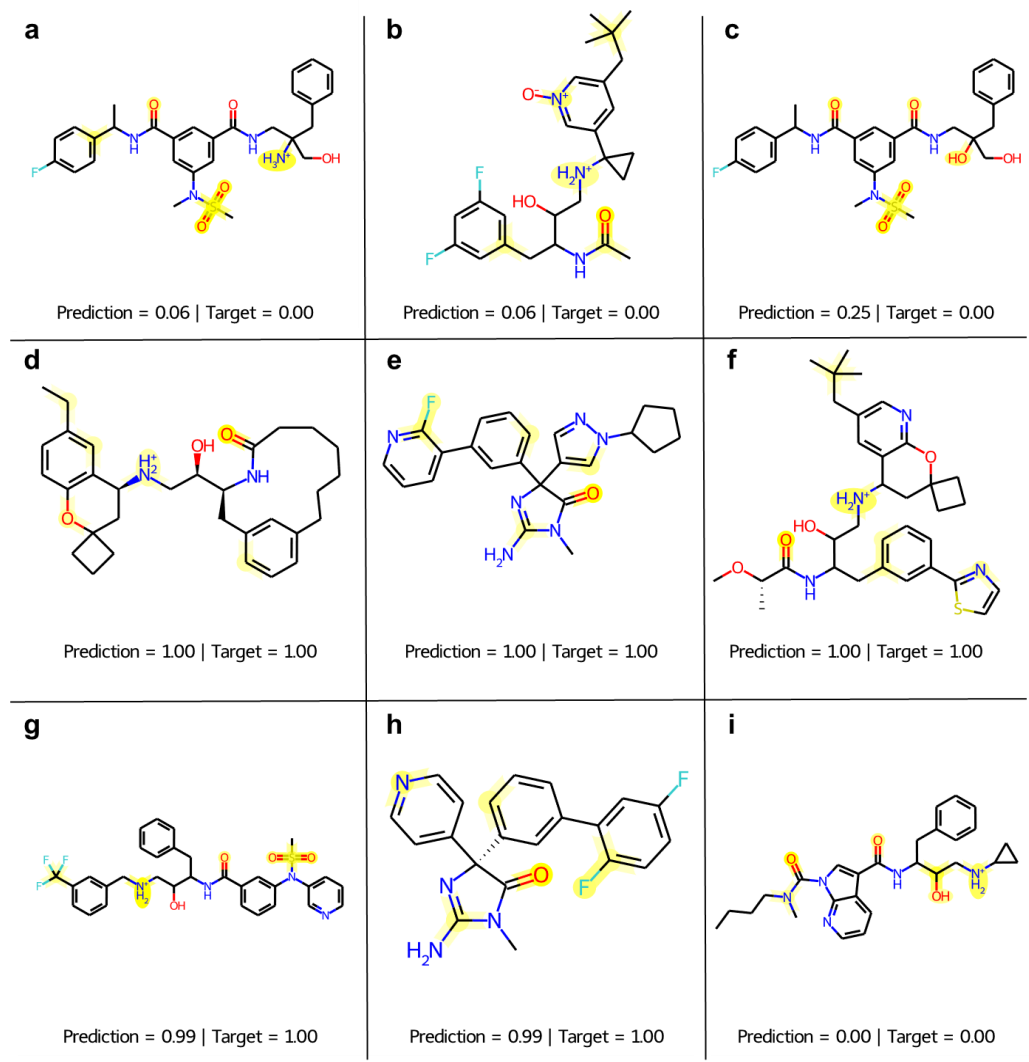

**Figure 4.** Attention-based interpretable analysis on BACE datasets. a-i illustrate assign different attention weight on different molecular substructures, which indicate the crucial substructures to the BACE target.

**Limitations**

Even though MolCAP achieves impressive performance via chemical reactivity knowledge and prompted finetuning, there are several limitations in our proposed method that deserve further research in the future.

**Dependencies on manual templet.** MolCAP relies on the manual templet to obtain high quality of performance and may incur thousands of trails and errors to search out the optimal prompting templet, which increases the computation cost and development duration. Therefore, the development of algorithm that can automated generate the optimal templet like some gradient-based methods[31,32] employed in the field of nature language processing is recommended.

**Information about the internal structure of molecules is not fully utilized.** In MolCAP, we merely focus on the effect brought by the chemical reactivity knowledge and neglect the abundant structural information including topology, geometry, and some motif-related information, limiting the improvement of final performance. So, the combined utilization of contrastive-base structure learning[16,33], the functional group discrimination[20], and the embedding of other forms of external knowledge such as element-wise knowledge graph[34] or geometries[17,35] are suggested to be employed in pretraining stage.

## Materials and Methods

**Multi-sense and multi-scale Graph Transformer (MSMS-GT) Encoder**

In order to reduce information loss in molecular representation learning and fit with our proposed pre-training framework, we used the MSMS-GT model proposed by Wang et al.[36], which has been proved to be an molecular representation model with informative bond embeddings. For better understanding, the corresponding nations are as follows. Given the molecular graph $\mathcal{G} = (V, A)$, the resulting representation $h_v^l$ of atom $v$ can be obtained through $l$ layers MSMS-GT encoder $f_{\theta^l}(\cdot)$, where $V \coloneqq \{v\}^N$ and $A$ denote the atom set with atom number of $N$ and the adjacent matrix of them, respectively. The iterative update process of $h_v^l$ can be generalized as follows:

$$h_v^l = f_{\theta^l}(\{h_u^{l-1}\}_{u \in \mathcal{V}}, A) \qquad (1)$$

where $\theta^l$ denotes the learnable parameters of $l$ layer.

**Pretraining**

**Molecular reactivity knowledge integrating.** To ensure the downstream tasks have acquired knowledge of molecular reactivity knowledge before the fine-tuning phase, we developed four predictive self-supervised agent tasks using chemical reaction datasets. More specifically, we encoded a set of input reactant molecules to predict the changes in their formal charge, hydrogen content, chirality, and bond order between atom pairs that occur after the given chemical reaction. Specifically, we formulated above four agent tasks as multi-classification tasks including three atom-level task and one bond-level task. The loss functions to optimize the learnable parameters is as follows:

$$\mathcal{L}_{att} = -\sum_{v \in V} \left( \log \frac{exp(p_{v;y^+;att})}{\sum_{y \in \mathcal{Y}} exp(p_{v;y})} \right),$$
$$att \in \{charge, hydrogen, chirality\}, \qquad (2)$$

$$\mathcal{L}_{bond} = -\sum_{(v,u) \in A} \left( \log \frac{exp(p_{vu;y^+;att})}{\sum_y exp(p_{vu;y})} \right), \qquad (3)$$

$$p_{v;att} = (h_v^l)^T W_{att} + B_{att}, \qquad (4)$$

$$p_{vu} = (h_v^l W_{bond} h_u^l)^T W'_{bond} + B_{bond} \qquad (5)$$

where $\mathcal{Y}$ denotes the total set of corresponding classes and $y^+$ is the class of ground truth.

**Balanced multi task learning.** Simultaneously optimizing multiple objectives is not an easy task, as it often involves conflicting issues resulting from discrepancies in the difficulty level of tasks and the magnitude of loss values. This problem can cause models to neglect those tasks that are difficult to optimize, thereby resulting in the loss of reactivity knowledge during the pre-training phase. To resolve this issue, we borrow the DAMT[36] framework for the reactivity knowledge learning. To provide a detailed explanation, we set a descent rate for the i-th loss $r_i^{(t)}$ in the t-th training step, which helps to gauge the complexity of the i-th task. Additionally, we introduce a normalizing coefficient $\alpha_i$ to standardize the magnitude of the i-th loss. By combining the above information, we can express the total loss of the t-th step, $\mathcal{L}_T^{(t)}$, as follows:

$$\mathcal{L}_T^{(t)} = \sum_{i=1}^{K_t} \left( \frac{exp\left(\frac{r_i^{(t)}}{\tau}\right)}{\sum_{j=1}^{K_t} exp\left(\frac{r_j^{(t)}}{\tau}\right)} \alpha_i^t \mathcal{L}_i^{(t)} \right), \quad r_i^{(t)} = \frac{\mathcal{L}_i^{(t-1)}}{\mathcal{L}_i^{(t-2)}}, \quad \alpha_i^t = \frac{n}{\sum_{j=t-n}^{t-n} \mathcal{L}_i^{(j)}}, \quad (5)$$

**Finetuning**

In addition to normal finetuning process that freezes pretrained parameters, we introduce two finetuning mode: manual prompt finetuning and auto-prompt finetuning. The former can bridge the gap between pretraining and finetuning; while the latter is to facilitate integrating other pretrained framework (such as geometric enhanced) with reactivity knowledge.

**Manual prompt fine-tuning.** Considering the inconsistency between the inputs of pre-training tasks (a set of reactants) and downstream fine-tuning tasks (a single drug molecule), which leads to the downstream tasks not being able to better utilize the knowledge learned from pre-training, we manually designed a set of reactant templates to bridge this gap. Specifically, we designed several small molecule templates, such as water molecules, acetone molecules, hydroxide ions, propane molecules, etc., and merged them before the drug input, thus simulating the input as a set of reactants rather than a single drug molecule.

**Auto-prompt fine-tuning.** Considering the transferability of pre-trained models for chemical reactions, we have provided a convenient interface for integrating chemical reaction knowledge into models based on other pre-training methods using the auto-prompt method. Specifically, let another pretrained model be $g_{\theta'}(\cdot)$ with learnable parameter $\theta'$ and the atom feature with bond feature that are pretrained by MolCAP be $\{h_v, v \in \mathcal{V}\}$ and $\{h_{uv}, (u,v) \in \mathcal{A}\}$. The resulting prompted molecular vectors can be:

$$h_{\theta';v}^l = g_{\theta'}^l(h_{\theta';v}^{l-1} * h_v, \{h_{\theta';uv}^{l-1} * h_{uv}, (u,v) \in \mathcal{A}\}), \quad (6)$$

where $l$ is the layer number to be considered, and $*$ denotes the element-wise multiplication operation.

## Dataset

**Pretraining.** We used reactions from USPTO granted patents collected by Lowe[37], which contains ~0.7 million chemical reactions after we cleaned up its data. We randomly sample 80% of the reactions for training, 10% for validation and the remaining for test.

**Self-supervised learning task settings.** We utilized node-level and edge-level tasks to pre-train MolCAP. We adopt a multi-meaning and multi-scale bond embedding strategy for molecular bonds and a topological embedding method for molecular atoms to capture important chemical information. The hidden molecular representations are learned from the two embedding items through a multi-head attention mechanism. Our pre-training consists of four tasks: Bond-order Change Prediction, Charge Change Prediction, Hydrogen number Change Prediction, and Chirality Change Prediction. We utilized a dynamic adaptive multi-task learning module, in which the obtained hidden molecular representations are fed into four deep neural decision units corresponding to the four pre-training tasks. By using the dynamic adaptive multi-task learning module, each subtask is trained equally.

**Finetuning on common benchmarks.** We evaluate MolCAP on six types of benchmark dataset with a total of thirteen datasets: (1) molecular targets—beta-secretase (BACE, a key target in Alzheimer's disease); (2) blood-brain barrier penetration (BBBP); (3) molecular toxicities—clinical trial toxicity (ClinTox) and toxicity using the Toxicology in the 21st Century (Tox21) and Toxicity Forecaster (ToxCast) databases; (4) the drug's metabolism and side effect resource (SIDER); (5) solubility—Free Solvation (FreeSolv) and Estimated Solubility (ESOL) and lipophilicity; (6) major metabolic enzymes for drug metabolism (CYP1A2, CYP2C9, CYP2C19, CYP2D6 and CYP3A4). We follow the same splitting method described in MoleculeNet. Both split methods can evaluate the generalization ability of the model on non-distributed data samples.

**Baselines.** We compare the proposed method with various competitive baselines. AttentiveFP is fingerprint-based model, SMILES-BERT, Mol2Vec, GROVER, N-GRAM are sequence-based models, and MPNN, DMPNN, MGCN, TrimNet, are graph-based models.

In addition, Chemception, ADMET-CNN and QSAR-CNN are molecular image-based representation models.

**Experimental Setup.** We split the training, verification and test sets of all datasets in a ratio of 8:1:1. Based on the pre-trained model on USPTO, we use several MSMS-GT sublayers to fine-tune each downstream task and output molecular embeddings. We use random scaffold split and found that MolCAP achieves SOTA results in predicting inhibitors versus non-inhibitors across all five major drug metabolism enzymes compared with three state-of-the-art molecular image-based representation models.

**Data and code availability**

The pre-training data used in our study are publicly known as USPTO-FULL dataset collected by Lowe[37], which is widely used for chemical prediction and retrosynthesis prediction; whereas the downstream benchmarks can be downloaded from MoleculeNet (https://moleculenet.org/datasets-1). In addition, the extended metabolism dataset can be found in TDC[38] website. The code of MolCAP can be found in https://github.com/wangyu-sd/MolCAP.

**Acknowledgements**

The authors acknowledge the anonymous reviewers for reviewing the manuscript.

**Author contributions**

L.W. and J. Z. conceived the basic idea and designed the research study. L.W. and J. Z. developed the method. Y.W., J.Z., and J.R. drew the figure. Y. W., J.Z. wrote the manuscript. L.W. reviewed the manuscript.

**Competing interests**

The authors declare that the research is conducted in the absence of any commercial or financial relationships that could be construed as a potential conflict of interest.

**Funding**

The work was supported by the Natural Science Foundation of China (Nos. 62071278 and 62072329) and Natural Science Foundation of Shandong Province (ZR2020ZD35).